\documentclass{article}
\usepackage{amsmath}
\usepackage{longtable}
\usepackage{graphicx}
\usepackage{amssymb}
\usepackage{amsmath}
\usepackage{graphicx}

\begin{document}

\begin{center}
\textbf{LENSING OBSERVABLES:\ MASSLESS DYONIC vis-\`{a}-vis ELLIS WORMHOLE }

\bigskip

R.F. Lukmanova$^{1,a}$, G.Y. Tuleganova$^{1,b}$, R.N. Izmailov$^{1,c}$,

and

K.K. Nandi$^{1,2,d}$

\bigskip

$^{1}$Zel'dovich International Center for Astrophysics, Bashkir State
Pedagogical University named after M.Akmullah, 3A, October Revolution Street, Ufa 450008, RB, Russia

$^{2}$High Energy and Cosmic Ray Research Center, University of North
Bengal, Siliguri 734 013, WB, India

\bigskip

$^{a}$E-mail: mira789@mail.ru

$^{b}$E-mail: gulira.tuleganova@yandex.ru

$^{c}$E-mail: izmailov.ramil@gmail.com

$^{d}$E-mail: kamalnandi1952@yahoo.co.in

-------------------------------------------------

\textbf{\bigskip }

\textbf{Abstract}
\end{center}

Stable massless wormholes are theoretically interesting in their own right
as well as for astrophysical applications, especially as galactic halo
objects. Therefore, the study of gravitational lensing observables for such
objects is of importance, and we do here by applying the
parametric post-Newtonian method of Keeton and Petters to massless
dyonic charged wormholes of the Einstein-Maxwell-Dilaton field theory
and to the massless Ellis wormhole of the Einstein minimally coupled scalar
field theory. The paper exemplifies how the lensing signatures of two
different solutions belonging to two different theories could be
qualitatively similar from the \textit{observational} point of view.
Quantitative differences appear depending on the parameter values.
Surprisingly, there appears an unexpected divergence in the correction to
differential time delay, which seems to call for a review of its original
derivation.

\begin{center}
\textbf{1. Introduction}
\end{center}

Gravitational lensing today is an inevitable part of astrophysicists'
toolkit for probing a number of interesting phenomena dealing from compact
objects to cosmology with widely varying distance scales. Especially, the
importance of studying lensing signatures in the weak field limit lies in
its ability to probe large-scale structures as well as the nature of the
lens (see, e.g., Ref [1]).The central role in the lensing is played by the
deflection of light caused by the gravitating lens, assumed here to be
static and spherically symmetric. Light deflection angles caused by several
Morris-Thorne traversable wormholes [2-8] and other objects [9-12] have been
studied in the strong and weak field limit. Among them, massless wormholes
are stable [13-15] and have received particular attention [16]. By
"massless", we mean that only the Keplerian mass is zero, while the energies
of other nonvanishing fields go into making what is called the "Wheelerian
mass" [17] that curves the space and it is this curvature that is revealed
by light deflection. The weak field light deflection angle has been recently
calculated in the literature by applying the Gauss-Bonnet Theorem (GBT)
[18-22] to a class of massless dyonic wormholes, re-interpreted also as the
Einstein-Rosen bridge.

Studying lensing signatures is a step farther than calculating merely the
deflection angle since the signatures take us into the realm of expected
observables. The motive of the present paper is to examine how these
observables differ for different massless lenses. For an important
application of lensing by the massless Ellis wormhole\footnote{%
To do justice, it should be called "Ellis-Bronnikov" wormhole since the two
authors Ellis [23] and Bronnikov [24] discovered the solution
independently and almost simultaneously in 1973. Nonetheless, we continue to
call it Ellis wormhole here in order to avoid confusion with the prevailing
nomenclature in the literature.}, which is a solution of
Einstein minimally coupled scalar (EMS) field theory, it was shown by Abe
[25] (this work was extended in Ref.[26]) that the weak field hypothesis is a
good approximation for lensing by our galaxy if the throat radius is less
than $10^{11}$ km. He further argued that if the massless wormholes, treated
as Galactic halo objects, are bound to the Galaxy with throat radii between
certain limits having a number density approximately equalling that of
ordinary stars, then their detection is possible by analyzing the past data.
Magnification of apparent brightness of distant stars lensed by an
intermediate wormhole is another important lensing effect. In view of these
astrophysically detectable effects and assuming that the halo region of
galaxies are populated by massless Ellis wormholes as conjectured by Abe
[25], it is of importance to calculate their observable signatures.

In this paper, we shall study the observables for massless dyonic wormhole
in the Einstein-Maxwell-dilaton (EMD)theory [27] vis-\`{a}-vis those of massless Ellis wormhole of the
EMS theory [23]. To that end, we shall first derive, using the
Keeton-Petters (KP) method [28-30], the weak field deflection and lensing
observables such as image positions, magnifications, centroid of a class of
spherically symmetric static massless dyonic wormholes in the EMD theory
that is already receiving attention\ (see, e.g., Ref. [16]). Next, we shall
calculate the same observables associated with the massless Ellis wormhole
in the EMS theory characterized by a scalar charge and tabulate a comparison
between the wormholes. All observables will be expressed as a function of
source angular position $\beta$. We shall graphically present quantitative
differences in lensing observables for the two wormholes and point out an
unexpected divergence in the correction to differential time delay.

The paper is section wise organized as follows. In Sec.II, we shall use the KP
method to verify the expression for the weak field deflection angle by
dyonic wormhole calculated by Jusufi \textit{et al} [16]. In Sec.III, we shall
calculate the lensing observables for the two massless wormholes under
consideration. Sec.IV summarizes the paper.

\begin{center}
\textbf{II. Weak field deflection angle by KP method}
\end{center}

Keeton and Petters [29] developed a very useful framework for computing
corrections to a core set of observable properties in a general
asymptotically flat metric theory of gravity. The focus is to demonstrate
how to handle lensing in competing gravity theories using
post-post-Newtonian (PPN) correction terms up to third-order. Their method
provides computation of \textit{observable }quantities that are essentially
coordinate independent and therefore are physically relevant. The readers
are urged to consult the original series of papers by the authors [29].

\bigskip The relevant EMD action is\footnote{%
Recently, Goulart [31] also derived phantom wormholes for a "sign reversed"
kinetic term $+2\partial _{\mu }\phi \partial ^{\mu }\phi $ in the EMD
action (1), but we are considering only dyonic wormholes here.} 
\begin{equation}
S_{\text{EMD}}=\int d^{4}x\sqrt{-g}\left[ R-2\partial _{\mu }\phi \partial
^{\mu }\phi -e^{-2\phi }F_{\mu \nu }F^{\mu \nu }\right] ,\text{ \ }F_{\mu
\nu }=\partial _{\mu }A_{\nu }-\partial _{\nu }A_{\mu }.
\end{equation}%
The dyonic massless wormhole in the ($t,r,\theta ,\varphi $) coordinates
derived by Goulart [27] and subsequently studied by Jusufi \textit{et al.}
[16] is

\begin{eqnarray}
d\tau ^{2} &=&-\frac{1}{1+\frac{a^{2}}{r^{2}}}dt^{2}+\frac{1+\frac{a^{2}}{%
r^{2}}}{1+\frac{k^{2}}{r^{2}}}dr^{2}+(r^{2}+a^{2})\left( d\theta ^{2}+\sin
^{2}\theta d\varphi ^{2}\right) , \\
a^{2} &=&2PQ\text{, }k^{2}=\Sigma ^{2}+a^{2},F_{rt}=Q/r^{2}\text{ }%
,F_{\theta \varphi }=P\sin \theta , \\
e^{2\phi } &=&e^{2\phi _{0}}\frac{r+d_{1}}{r+d_{0}},\text{ }%
d_{1}=-d_{0}=-\Sigma .
\end{eqnarray}%
The solution represents a three-parameter wormhole characterized by electric
charge ($Q$), magnetic charge ($P$) and a dilatonic charge ($\Sigma $). For
the special case $\Sigma =0$, this solution exactly coincides with the
Einstein-Rosen bridge [22]. To apply the KP method, we express the metric
(2)\ in isotropic coordinates ($t,R,\theta ,\varphi $) by introducing the
transformation 
\begin{equation}
r=\frac{R^{2}-k^{2}}{2R},
\end{equation}%
which when inverted yields $R=\frac{1}{2}r\left( 1\pm \sqrt{1+k^{2}/r^{2}}%
\right) $. Discarding the negative sign, we find $\frac{R}{r}\rightarrow 1$
as $r\rightarrow \infty $, so at large distances $R$ and $r$ coincide. The
metric (2) under the radial transformation (5) becomes%
\begin{eqnarray}
d\tau ^{2} &=&-A(R)dt^{2}+B(R)\left( dR^{2}+R^{2}d\theta ^{2}+R^{2}\sin
^{2}\theta d\varphi ^{2}\right)   \notag \\
&=&-\left[ \frac{1}{1+\frac{4R^{2}a^{2}}{\left( R^{2}-k^{2}\right) ^{2}}}%
\right] dt^{2}  \notag \\
&&+\left[ \frac{k^{4}+4a^{2}R^{2}-2k^{2}R^{2}+R^{4}}{4R^{4}}\right] \left(
dR^{2}+R^{2}d\theta ^{2}+R^{2}\sin ^{2}\theta d\varphi ^{2}\right) .
\end{eqnarray}%
This is an asymptotically flat metric, which is invariant under inversion: $%
R\rightarrow \frac{k^{2}}{R}$. The metric (2) thus represents a twice
asymptotically flat regular wormhole as the spacetimes on either side of the
throat appearing at $r_{\text{th}}=a=\sqrt{2PQ}$ (minimum areal radius) or
at the isotropic radius $R_{\text{th}}=\frac{1}{2}a\left( 1+\sqrt{%
1+k^{2}/a^{2}}\right) $ are regular. The tidal forces can also be verified
to be finite everywhere. Now redefine $R=2\overline{R}$ so that 
\begin{equation}
d\tau ^{2}=-A(\overline{R})dt^{2}+B(\overline{R})\left( d\overline{R}^{2}+%
\overline{R}^{2}d\theta ^{2}+\overline{R}^{2}\sin ^{2}\theta d\varphi
^{2}\right) 
\end{equation}%
and the metric functions expand as%
\begin{equation}
A(\overline{R})=1-\frac{a^{2}}{\overline{R}^{2}}-\frac{a^{4}}{\overline{R}%
^{4}}\left( \frac{k^{2}}{2a^{2}}\right) +...
\end{equation}

\begin{equation}
B(\overline{R})=1+\frac{a^{2}}{\overline{R}^{2}}\left( 1-\frac{k^{2}}{2a^{2}}%
\right) +\frac{a^{4}}{\overline{R}^{4}}\left( \frac{k^{4}}{16a^{4}}\right)
+...
\end{equation}%
Following the method of Keeton and Petters [29], and taking the PPN
potential to be\footnote{%
Here $a$ is the so-called "Wheelerian mass" made of the electric and
magnetic field energies, while the dilaton $\Sigma $ does not manifestly
contribute to the potential.}%
\begin{equation}
\frac{\Phi }{c^{2}}=\frac{a}{\overline{R}},
\end{equation}%
we can have a PPN expansion as%
\begin{align}
A(\overline{R})& =1+2\alpha ^{\prime }\left( \frac{\Phi }{c^{2}}\right)
+2\beta ^{\prime }\left( \frac{\Phi }{c^{2}}\right) ^{2}+\frac{3}{2}\xi
^{\prime }\left( \frac{\Phi }{c^{2}}\right) ^{3}+... \\
B(\overline{R})& =1-2\gamma ^{\prime }\left( \frac{\Phi }{c^{2}}\right) +%
\frac{3}{2}\delta ^{\prime }\left( \frac{\Phi }{c^{2}}\right) ^{2}-\frac{1}{2%
}\eta ^{\prime }\left( \frac{\Phi }{c^{2}}\right) ^{3}+...
\end{align}%
Since the KP method is relatively new, from here on, we outline the steps
connecting the above coefficients $\alpha ^{\prime },\beta ^{\prime },\gamma
^{\prime }$ etc., for the isotropic form to the new coefficients $%
a_{1},b_{1},a_{2},b_{2},a_{3},b_{3}$ for the standard form and then to the
final coefficients $A_{1},A_{2},A_{3}$. The latter coefficients all relate
to the PPN expansion of the metric (7) written in the \textit{standard}
coordinates ($t,\rho ,\theta ,\varphi $) in the form (in general, coordinate
choices do not change physics, but in the standard coordinate system the
surface area of a sphere is given by the familiar expression $4\pi \rho ^{2}$%
) 
\begin{equation}
d\tau ^{2}=-f(\rho )dt^{2}+g(\rho )d\rho ^{2}+\rho ^{2}\left( d\theta
^{2}+\sin ^{2}\theta d\varphi ^{2}\right) ,
\end{equation}%
where%
\begin{equation}
f(\rho )=A(\overline{R}),g(\rho )d\rho ^{2}=B(\overline{R})d\overline{R}%
^{2},\rho ^{2}=B(\overline{R})\overline{R}^{2}
\end{equation}%
and the corresponding potential will be $\left( \frac{a}{\rho }\right) $.
These transformations determine $f(\rho )$, $g(\rho )$ and connect the
coefficients as desired. The next step is to note that the impact parameter $%
b$ is related to the closest approach distance $\rho _{0}$ by 
\begin{equation}
\frac{1}{b^{2}}=\frac{f(\rho _{0})}{\rho _{0}^{2}},
\end{equation}%
which allows one to obtain 
\begin{equation}
\rho _{0}=b\left[ 1-a_{1}\left( \frac{a}{b}\right) +\frac{2a_{2}-3a_{1}^{2}}{%
2}\left( \frac{a}{b}\right) ^{2}+...\right] .
\end{equation}%
The final step is to expand the integrand in the exact deflection angle [34]
\begin{equation*}
\widehat{\alpha }(\rho _{0})=2\int_{\rho _{0}}^{\infty }\frac{1}{\rho ^{2}}%
\sqrt{\frac{f(\rho )g(\rho )}{1/b^{2}-f(\rho )/\rho ^{2}}}d\rho -\pi 
\end{equation*}%
in terms of the small PPN parameter $h=\frac{a}{\rho _{0}}$, integrate term
by term, express $\rho _{0}$ in terms of $b$, which would immediately yield
the coefficients $A_{1},A_{2},A_{3}$ of (22)-(24). By comparing similar
powers between (8), (11) and between (9), (12), one finds 
\begin{eqnarray}
a_{1} &=&\alpha ^{\prime }=0\text{, }b_{1}=\gamma ^{\prime }=0\text{, }%
a_{2}=\beta ^{\prime }-\alpha ^{\prime }\gamma ^{\prime }=-\frac{1}{2}, \\
b_{2} &=&\frac{3\delta ^{\prime }+\gamma ^{\prime 2}}{4}=\frac{1}{2}\left( 1-%
\frac{k^{2}}{2a^{2}}\right) \text{, }\xi ^{\prime }=0\text{, }\eta ^{\prime
}=0, \\
a_{3} &=&\frac{3\xi ^{\prime }+3\alpha ^{\prime }\delta ^{\prime }-8\beta
^{\prime }\gamma ^{\prime }+2\alpha ^{\prime }\gamma ^{\prime 2}}{4}=0, \\
b_{3} &=&\frac{3\eta ^{\prime }+15\delta ^{\prime }\gamma ^{\prime }-2\gamma
^{\prime 3}}{16}=0,
\end{eqnarray}%
The two way deflection angle is 
\begin{equation}
\widehat{\alpha }(b)=A_{1}\left( \frac{a}{b}\right) +A_{2}\left( \frac{a}{b}%
\right) ^{2}+A_{3}\left( \frac{a}{b}\right) ^{3}+...
\end{equation}%
where $b$ is the invariant impact parameter related to the closest approach
distance $r_{0}$ to leading order by $b=r_{0}\left( 1+a_{1}\frac{a}{%
\overline{R}}\right) =r_{0}$. Also%
\begin{eqnarray}
A_{1} &=&2(a_{1}+b_{1}), \\
A_{2} &=&\left( 2a_{1}^{2}-a_{2}+a_{1}b_{1}-\frac{b_{1}^{2}}{4}+b_{2}\right)
\pi , \\
A_{3} &=&\frac{2}{3}%
[35a_{1}^{3}+15a_{1}^{2}b_{1}-3a_{1}(10a_{2}+b_{1}^{2}-4b_{2})  \notag \\
&&+6a_{3}+b_{1}^{3}-6a_{2}b_{1}-4b_{1}b_{2}+8b_{3}].
\end{eqnarray}%
In view of the values in (17)-(20), we find%
\begin{equation}
A_{1}=0\text{, }A_{2}=\left( 1-\frac{a^{2}+\Sigma ^{2}}{4a^{2}}\right) \pi 
\text{, }A_{3}=0,
\end{equation}%
so the invariant deflection angle is%
\begin{equation}
\widehat{\alpha }(b)=A_{2}\left( \frac{a}{b}\right) ^{2}=\frac{3\pi PQ}{%
2b^{2}}-\frac{\pi \Sigma ^{2}}{4b^{2}}.
\end{equation}

This result exactly agrees with the deflection angle\ obtained by Jusufi 
\textit{et al.} [16], who used the GBT method. This agreement suggests that
the potential $\frac{\Phi }{c^{2}}=\frac{a}{\overline{R}}$ is the correct
one, which does not contain the dilatonic charge $\Sigma $, although
interestingly it does contribute to the deflection angle and the actual
lensing observables. Incidentally, since $A_{3}=0$, the third order term $%
\left( \frac{a}{b}\right) ^{3}$ is \textit{absent} in the deflection. This
non-trivial information\ about the weak field deflection is difficult to
obtain by GBT method but here it is easily obtained from the Keeton-Petters
method.

The weak field deflection $\widehat{\alpha }$ in general has a major
difference with strong field deflection. The strong field deflection
suffered by light rays passing at an invariant impact parameter $b$ closest
to the photon sphere have a logarithmic divergence [32,33]. This fact
prevents the \textit{exact} deflection angle to be Taylor expanded to yield
the same light deflection for the same $b$. For instance [33], for the
Schwarzschild black hole of mass $M$,

\begin{equation}
\widehat{\alpha }_{\text{strong}}(b^{\prime })=-\pi +\log \left[ \frac{%
216(7-4\sqrt{3})}{b^{\prime }}\right] +O(b^{\prime }),
\end{equation}%
\begin{equation}
\widehat{\alpha }_{\text{weak}}(b^{\prime })=\frac{4}{3\sqrt{3}}\left(
1-b^{\prime })+O(1-b^{\prime }\right) ^{2},
\end{equation}%
where the redefined common impact parameter $b^{\prime }$ is\ $1-b^{\prime }=%
\frac{3\sqrt{3}M}{b}$. When $b=3\sqrt{3}M$, $\widehat{\alpha }_{\text{strong}%
}\rightarrow \infty $, but $\widehat{\alpha }_{\text{weak}}=\frac{4M}{b}$,
as expected. These facts indicate that the weak field lensing is expected to
yield a set of lensing observables completely different from those of the
strong field. We note that $\widehat{\alpha }_{\text{strong}}$ is itself an
approximation in the strong regime with $O(b^{\prime })$ neglected.\footnote{%
We thank an anonymous referee for pointing it out.}

\begin{center}
\textbf{III. Lensing observables}
\end{center}

\begin{figure}[h]
\centerline { \includegraphics 
[scale=0.3]{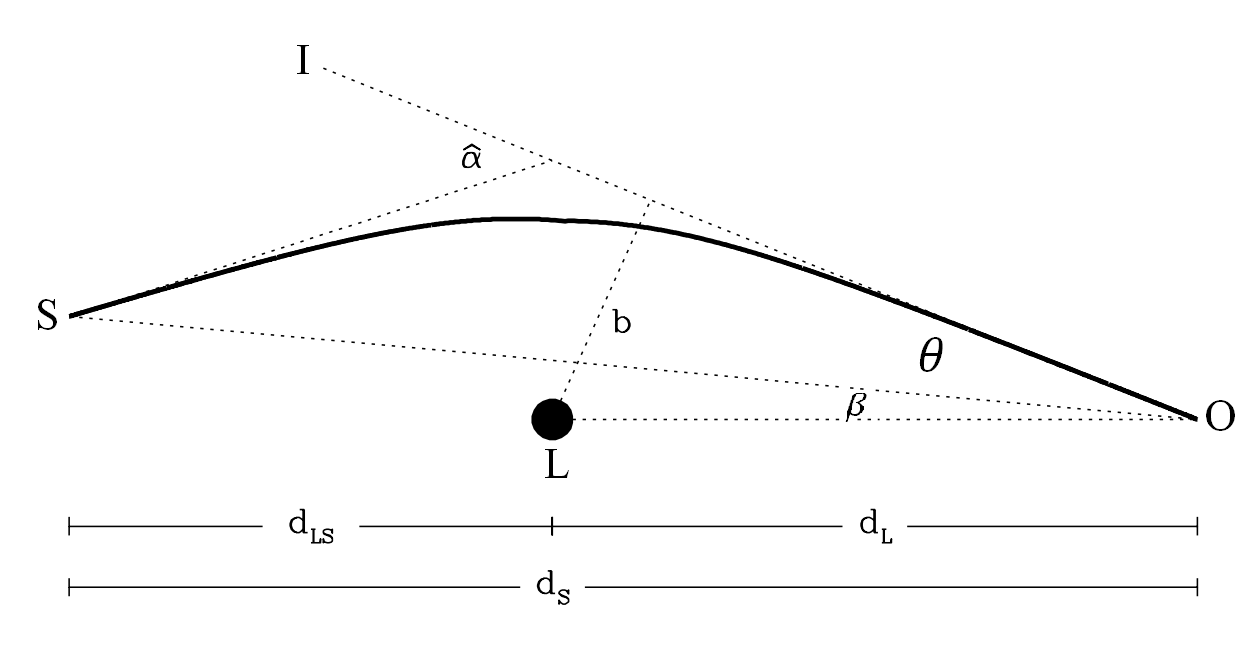}}
\caption{Lens geometry.}
\end{figure}

The lens geometry in shown in Fig.1. The corresponding lens equation, with
the angles scaled by the Einstein angle $\theta _{E}$, is [35]
\begin{equation}
\tan \beta =\tan \theta -D\ \left[ \tan \theta +\tan (\hat{\alpha}-\theta )%
\right] ,
\end{equation}%
where $\hat{\alpha}$ is the light deflection angle, $\beta $ is the source
angular position, $\theta $ is the angular position of the image and $%
D\equiv d_{LS}/d_{S}$. This equation, though obtained from elementary
trigonometry of Fig.1, very well describes the full relativistic treatment
for light propagation. The next step is to expand the angular position $%
\theta $ of the image as%
\begin{equation}
\theta =\theta _{0}+\theta _{1}\varepsilon +\theta _{2}\varepsilon
^{2}+O(\varepsilon ^{3}),
\end{equation}%
where $\theta _{0}$ represents the image position in the weak deflection
limit, while $\theta _{1}$, $\theta _{2}$ represent first- and second-order
correction terms, $\theta _{E}$ is the Einstein angle, $\varepsilon $ is the
small perturbative parameter defined by 
\begin{equation}
\varepsilon \equiv \frac{\theta _{E}}{4D},\text{ }\theta _{E}\equiv \sqrt{%
\frac{4Gad_{LS}}{c^{2}d_{L}d_{S}}}.
\end{equation}%
Using the expansion for $\theta $, the exact bending angle can be expanded
as [31] (terms in $\varepsilon ^{3}$ not shown, to save space) 
\begin{equation}
\widehat{\alpha }=\frac{A_{1}}{\theta _{0}}\varepsilon +\frac{%
A_{2}-A_{1}\theta _{1}}{\theta _{0}^{2}}\varepsilon ^{2}+...
\end{equation}%
Substituting $\theta $ and $\widehat{\alpha }$ in the lens equation and
expanding it beyond linear order, one has%
\begin{equation}
0=D\left[ -4\beta +4\theta _{0}-\frac{A_{1}}{\theta _{0}}\right] \varepsilon
+\frac{D}{\theta _{0}^{2}}\left[ -A_{2}+\left( A_{1}+4\theta _{0}^{2}\right)
\theta _{1}\right] \varepsilon ^{2}+...
\end{equation}%
The lensing observables are obtained as follows. Note that each coefficient
in the expression (33) should vanish since $\varepsilon $ can be
independently varied, which then yields the first observable, the\textit{\
image position} $\theta _{0}$ and correction $\theta _{1}$ to it (the
expression for $\theta _{2}$ not shown)%
\begin{equation}
\theta _{0}=\frac{1}{2}\left[ \sqrt{A_{1}+\beta ^{2}}+\beta \right] \text{,
\ }\theta _{1}=\frac{A_{2}}{A_{1}+4\theta _{0}^{2}}.
\end{equation}

\textit{Magnification} $\mu \left( \theta \right) $ of images takes place
because the bending of light by the lens focusses more light rays from the
source into a solid angle at the observer brightening up the image. It is
defined by%
\begin{equation}
\mu \left( \theta \right) =\left[ \frac{\sin \beta }{\sin \theta }\frac{%
d\beta }{d\theta }\right] ^{-1}.
\end{equation}%
Exactly the same procedure as for the image position and correction applies
to this case too. By using the expressions for $\theta _{1},\theta _{2}$,
one can write a series expansion as 
\begin{equation}
\mu =\mu _{0}+\mu _{1}\varepsilon +\mu _{2}\varepsilon ^{2}+O(\varepsilon
)^{3},
\end{equation}%
which yields magnification $\mu _{0}$ and its correction $\mu _{1}$ ($\mu
_{2}$ not shown)%
\begin{equation}
\mu _{0}=\frac{16\theta _{0}^{4}}{16\theta _{0}^{4}-A_{1}^{2}}\text{, \ }\mu
_{1}=\frac{16A_{2}\theta _{0}^{3}}{\left( A_{1}+4\theta _{0}^{2}\right) ^{3}}%
.
\end{equation}

In the case when the individual images are too close together and cannot be
resolved, it is useful to define \textit{total magnification} $\mu _{\text{%
tot}}$ as a sum of magnification of positive and negative parity images $\mu
^{+}$ and $\mu ^{-}$ as%
\begin{eqnarray}
\mu _{\text{tot}} &=&\left\vert \mu ^{+}\right\vert +\left\vert \mu
^{-}\right\vert  \notag \\
&=&\frac{16A_{1}^{2}\left( \theta _{0}^{8}-1\right) }{\left( 16\theta
_{0}^{4}-A_{1}^{2}\right) \left( A_{1}^{2}\theta _{0}^{2}-16\right) }  \notag
\\
&&-\frac{16\left( A_{1}-4\right) A_{2}\theta _{0}^{3}}{\left( A_{1}+4\theta
_{0}^{2}\right) ^{3}\left( 4+A_{1}\theta _{0}^{2}\right) ^{3}}\{\left[
16+A_{1}\left( 4+A_{1}\right) \right]  \notag \\
&&\times \left( \theta _{0}^{6}-1\right) +12A_{1}\theta _{0}^{2}\left(
\theta _{0}^{2}-1\right) \}{\varepsilon +}O\left( {\varepsilon }\right) ^{2}.
\end{eqnarray}%
The center of light or in short the \textit{centroid} $\Theta _{\text{cent}}$
of the images is simply the magnification-weighted sum of the image
positions and its expansion is 
\begin{equation}
\Theta _{\text{cent}}=\Theta _{\text{cent,}0}+\Theta _{\text{cent,}1}{%
\varepsilon }+\Theta _{\text{cent,}2}{\varepsilon }^{2}{+}O\left( {%
\varepsilon }\right) ^{3}
\end{equation}

\begin{equation}
\Theta _{\text{cent,}0}=\left\vert \beta \right\vert \frac{3A_{1}+4\beta ^{2}%
}{2A_{1}+4\beta ^{2}},\text{ }\Theta _{\text{cent,}1}=0,
\end{equation}%
where $\Theta _{\text{cent,}2}$ is not shown.

The \textit{differential time delay} $\Delta \hat{\tau}$ is the delay in the
arrival times at the observer from a pair of images and is also PPN expanded
in terms of $\varepsilon $ (see below).

\textbf{(A) Dyonic massless wormhole}

The PPN parameters in standard coordinates in this case are 
\begin{equation}
a_{1}=a_{3}=b_{1}=b_{3}=0,\;a_{2}=-\frac{1}{2},\;b_{2}=\frac{1}{2}-\frac{%
k^{2}}{4a^{2}},
\end{equation}
\begin{equation}
A_{1}=0,\;A_{2}=\pi \left( 1-\frac{k^{2}}{4a^{2}}\right) ,\;A_{3}=0.
\end{equation}
The lensing observables are as follows.

(i) Image position and corrections 
\begin{equation}
\theta _{0}=\beta ,\;\theta _{1}=\frac{\pi }{4\beta ^{2}}\left( 1-\frac{k^{2}%
}{4a^{2}}\right) ,\;\theta _{2}=-\frac{\pi ^{2}}{8\beta ^{5}}\left( 1-\frac{%
k^{2}}{4a^{2}}\right) ^{2}.
\end{equation}

(ii) Magnification and corrections
\begin{equation}
\mu _{0}=1,\;\mu _{1}=-\frac{\pi }{4\beta ^{3}}\left( 1-\frac{k^{2}}{4a^{2}}%
\right) ,\;\mu _{2}=-\frac{3\pi ^{2}}{8\beta ^{6}}\left( 1-\frac{k^{2}}{%
4a^{2}}\right) ^{2}.
\end{equation}

(iii) Total magnification, centroid and corrections 
\begin{equation}
\mu _{\text{tot}}=\frac{3\pi ^{2}}{4\beta ^{6}}\left( 1-\frac{k^{2}}{4a^{2}}%
\right) \varepsilon ^{2},
\end{equation}%
\begin{equation}
\Theta _{\text{cent},0}=|\beta |,\;\Theta _{\text{cent},1}=0,\;\Theta _{%
\text{cent},2}=-\frac{3\pi ^{2}}{8\beta ^{5}}\left( 1-\frac{k^{2}}{4a^{2}}%
\right) ^{2}.
\end{equation}

(iv) Differential time delay and corrections 
\begin{equation}
\Delta \hat{\tau}=\Delta \hat{\tau}_{0}\ +\ \Delta \hat{\tau}%
_{1}\,\varepsilon \ +\ O\left( {\varepsilon }\right) ^{2},
\end{equation}%
where 
\begin{eqnarray}
\Delta \hat{\tau}_{0} &=&\frac{1}{2}\,|\beta |\sqrt{A_{1}+\beta ^{2}}\ +\ 
\frac{A_{1}}{4}\ \ln \left( \frac{\sqrt{A_{1}+\beta ^{2}}+\beta }{\sqrt{%
A_{1}+\beta ^{2}}-\beta }\right) \,, \\
\Delta \hat{\tau}_{1} &=&\frac{A_{2}}{A_{1}}\ |\beta |\,.
\end{eqnarray}%
In the present case, since $A_{1}=0$, we find 
\begin{equation}
\Delta \hat{\tau}_{0}=\frac{\beta ^{2}}{2},\;\Delta \hat{\tau}%
_{1}\rightarrow \text{divergent}.
\end{equation}

One way to interpret this divergence in the differential time delay
correction is the following. If we set $A_{2}=0$, $\beta \neq 0$, then the
correction $\Delta \hat{\tau}_{1}$ vanish and measurement of $\Delta \hat{%
\tau}_{0}$ would allow one to determine location of the angular position $%
\beta $ of the source. When $A_{2}\neq 0$, $\beta =0$, the
source, lens, and observer are aligned and in this case, one has an \textit{%
Einstein ring} from where light reaches the observer exactly at the same
time so that there is no time delay, hence, $\Delta \hat{\tau}_{0}=0$. Since
there are no individual images now, the corrections attributable to
individual images also lose their meaning, a symptom of which is the
appearance of divergences in Eqs.(50).

\textbf{(b) Ellis massless wormhole}

The action is 
\begin{equation}
S_{\text{EMS}}=\int d^{4}x\sqrt{-g}\left[ R+2\partial _{\mu }\Psi \partial
^{\mu }\Psi \right] ,
\end{equation}%
where the kinetic term $+2\partial _{\mu }\Psi \partial ^{\mu }\Psi $ is
sign reversed here compared to that in action (1) meaning that the field $%
\Psi $ represents exotic phantom matter. The Ellis massless solution is
given by%
\begin{eqnarray}
d\tau ^{2} &=&-dt^{2}+d\ell ^{2}+(\ell ^{2}+m^{2})\left( d\theta ^{2}+\sin
^{2}\theta d\varphi ^{2}\right) , \\
\Psi &=&\frac{1}{\sqrt{2}}\left[ \frac{\pi }{2}-2\tan ^{-1}\left( \frac{\ell 
}{m}\right) \right] ,
\end{eqnarray}%
where $m$ is a constant of integration that can be called the scalar charge
proportional to the integrated total energy of the scalar field $\Psi $.
Under the transformation $\ell ^{2}+m^{2}=\rho ^{2}$, the metric reduces in
standard coordinates ($t,\rho ,\theta ,\varphi $) to the form%
\begin{equation}
d\tau ^{2}=-dt^{2}+\frac{d\rho ^{2}}{1-\frac{m^{2}}{\rho ^{2}}}+\rho
^{2}\left( d\theta ^{2}+\sin ^{2}\theta d\varphi ^{2}\right) .
\end{equation}%
Taking the PPN potential $\frac{\psi }{c^{2}}=\frac{m}{\rho }$, we find the
coefficients to be

\begin{equation}
a_{1}=a_{2}=a_{3}=b_{1}=b_{3}=0,\;b_{2}=\frac{1}{4},
\end{equation}%
\begin{equation}
A_{1}=0,\;A_{2}=\frac{\pi }{4},\;A_{3}=0.
\end{equation}%
The bending angle then follows as 
\begin{equation}
\widehat{\alpha }_{\text{weak}}(b)=\frac{\pi m^{2}}{4b^{2}},
\end{equation}%
which exactly reproduces the leading order term of the deflection calculated
by Bhattacharya and Potapov [6] by three independent ways other than the KP
method.

(i) Image position and corrections:

\begin{equation}
\theta _{0}=\beta ,\;\theta _{1}=\frac{\pi }{16\beta ^{2}},\;\theta _{2}=-%
\frac{\pi ^{2}}{128\beta ^{5}}.
\end{equation}

(ii) Magnification and corrections:

\begin{equation}
\mu _{0}=1,\;\mu _{1}=-\frac{\pi }{16\beta ^{3}},\;\mu _{2}=-\frac{3\pi ^{2}%
}{128\beta ^{6}}.
\end{equation}

(iii) Total magnification, centroid and corrections:

\begin{equation}
\mu _{\text{tot}}=\frac{3\pi ^{2}\varepsilon ^{2}}{64\beta ^{6}},
\end{equation}

\begin{equation}
\Theta _{\text{cent},0}=|\beta |,\;\Theta _{\text{cent},1}=0,\;\Theta _{%
\text{cent},2}=-\frac{3\pi ^{2}}{128\beta ^{5}}.
\end{equation}

(iv) Differential time delay and corrections:

The same expressions as in (48) and (49) apply, so that

\begin{equation}
\Delta \hat{\tau}_{0}=\frac{\beta ^{2}}{2},\;\Delta \hat{\tau}%
_{1}\rightarrow \text{ divergent.}
\end{equation}

Same arguments about divergence following Eq.(50) apply here too and need
not be repeated. The foregoing results are tabulated in Table 1 below for easy
comparison.

\begin{center}
\textbf{TABLE I }

\textbf{PPN observables for dyonic and Ellis wormholes }

\begin{tabular}{|l|l|l|}
\hline
Observable & Dyonic wormhole & Ellis wormhole \\ \hline
$\theta _{0}$ & $\beta $ & $\beta $ \\ \hline
$\theta _{1}$ & $\frac{\pi }{4\beta ^{2}}\left( 1-\frac{k^{2}}{4a^{2}}%
\right) $ & $\frac{\pi }{16\beta ^{2}}$ \\ \hline
$\theta _{2}$ & $-\frac{\pi ^{2}}{8\beta ^{5}}\left( 1-\frac{k^{2}}{4a^{2}}%
\right) ^{2}$ & $-\frac{\pi ^{2}}{128\beta ^{5}}$ \\ \hline
$\mu _{0}$ & $1$ & $1$ \\ \hline
$\mu _{1}$ & $-\frac{\pi }{4\beta ^{3}}\left( 1-\frac{k^{2}}{4a^{2}}\right) $
& $-\frac{\pi }{16\beta ^{3}}$ \\ \hline
$\mu _{2}$ & $-\frac{3\pi ^{2}}{8\beta ^{6}}\left( 1-\frac{k^{2}}{4a^{2}}%
\right) ^{2}$ & $-\frac{3\pi ^{2}}{128\beta ^{6}}$ \\ \hline
$\mu _{\text{tot}}$ & $\frac{3\pi ^{2}}{4\beta ^{6}}\left( 1-\frac{k^{2}}{%
4a^{2}}\right) \varepsilon ^{2}$ & $\frac{3\pi ^{2}}{64\beta ^{6}}%
\varepsilon ^{2}$ \\ \hline
$\Theta _{\text{cent},0}$ & $|\beta |$ & $|\beta |$ \\ \hline
$\Theta _{\text{cent},1}$ & $0$ & $0$ \\ \hline
$\Theta _{\text{cent},2}$ & $-\frac{3\pi ^{2}}{8\beta ^{5}}\left( 1-\frac{%
k^{2}}{4a^{2}}\right) ^{2}$ & $-\frac{3\pi ^{2}}{128\beta ^{5}}$ \\ \hline
$\Delta \hat{\tau}_{0}$ & $\frac{\beta ^{2}}{2}$ & $\frac{\beta ^{2}}{2}$ \\ 
\hline
$\Delta \hat{\tau}_{1}$ & divergent & divergent \\ \hline
\end{tabular}

\bigskip

\textbf{4. Summary}
\end{center}

The purpose of this paper was to investigate gravitational lensing
signatures of massless asymptotically flat wormholes in the two theories
described by the two actions (1) and (51). It is evident that the kinetic
term has different signs indicating that the nature of the source matter is
quite different in either action. The energy conditions are violated at the
solution level necessary to make the two solutions wormholes. Further, the
metrics do not coincide in an ordinary one-to-one correspondence of their
parameters on a real line, and neither are they connected by any coordinate
transformation. So the solutions are non-trivially different. It turns out
that the Ellis metric (52) follows from the dyonic metric (2) only when $a=0$%
, and $\Sigma ^{2}=-m^{2}$, meaning an \textit{imaginary} dilatonic charge.
It shows that the two metrics represent physically different wormholes. On
the other hand, the dilaton $\Sigma $ does not contribute to the central
potential (10) though it does contribute to energy conditions and
observables. The situation therefore is a very curious one deserving a
closer scrutiny of the lensing behavior of the two objects, which we have
done above.

We computed weak field lensing observables by applying the PPN method of
Keeton and Petters (KP) to massless dyonic and Ellis wormholes. The paper
nicely exemplifies how the observable lensing signatures of two physically
different objects originating from very different parent theories could
still be qualitatively similar from the \textit{observational} point of
view. Quantitative differences appear depending on the parameter values.
Quantitative differences appear depending on the parameter values $P$, $Q$, $%
\Sigma $ and on the source angular position $\beta $. In the special case $%
\Sigma =\sqrt{6PQ}$, where $P$ and $Q$ are magnetic and electric charge
respectively, the observables between the two wormholes drastically differ -
all the correction factors vanish for the dyonic charged wormhole, while
they remain nonzero for the Ellis wormhole. The observables in the two
cases are tabulated for easy view. For illustrative purposes and numerical
comparison, we take the black hole SgrA* residing at the center of our
galaxy and treat it as a massless wormhole made by a high concentration of
"Wheelerian mass" (made of $P,Q$ or $m$). In that case, it follows that [27] 
$\varepsilon =1.3\times 10^{-4}\left( \frac{d_{\text{LS}}}{10\text{ pc}}%
\right) ^{-1/2}$, which is used for all the Figs.2-7.

We notice that the correction $\Delta \hat{\tau}_{1}$ to differential time
delay from individual images surprisingly \textit{diverge} in both the
wormholes since $A_{1}=0$ despite the fact that all other observables are
finite as tabulated. Such a divergence in the weak field is thus an
unexpected behavior. Probably, it signals that the derivation of $\Delta 
\hat{\tau}_{1}$ needs to be reviewed. The divergence can be avoided only if $%
\beta =0$, which means the source, lens and the observer are to be situated
on a straight line and the image will be an Einstein ring instead of
individual images. There would be no differential time delay in this case.

\begin{figure}[b]
\centerline { \includegraphics 
[scale=0.6]{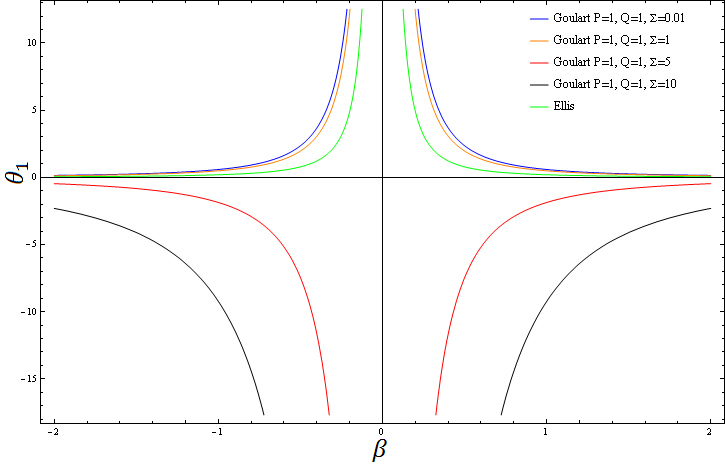}}
\caption{The second-order correction $\protect\theta _{1}$ for the source
angular position $\protect\beta \in $ $[-2;2].$}
\end{figure}

\begin{figure}[h]
\centerline { \includegraphics 
[scale=0.6]{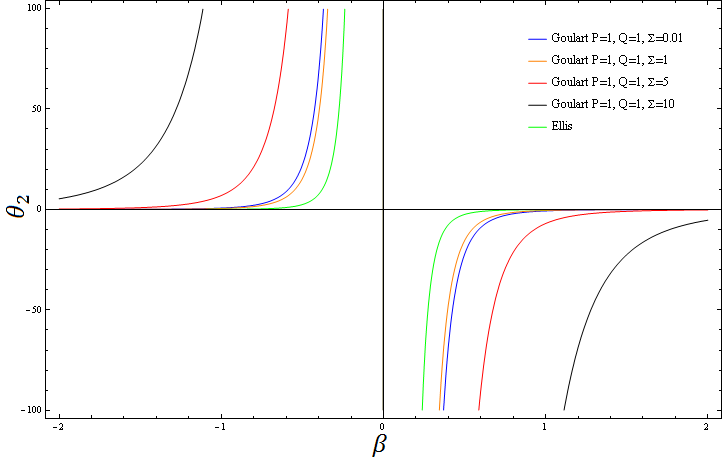}}
\caption{The second-order correction $\protect\theta _{2}$ for the source
angular position $\protect\beta \in $ $[-2;2].$}
\end{figure}

\begin{figure}[h]
\centerline { \includegraphics 
[scale=0.6]{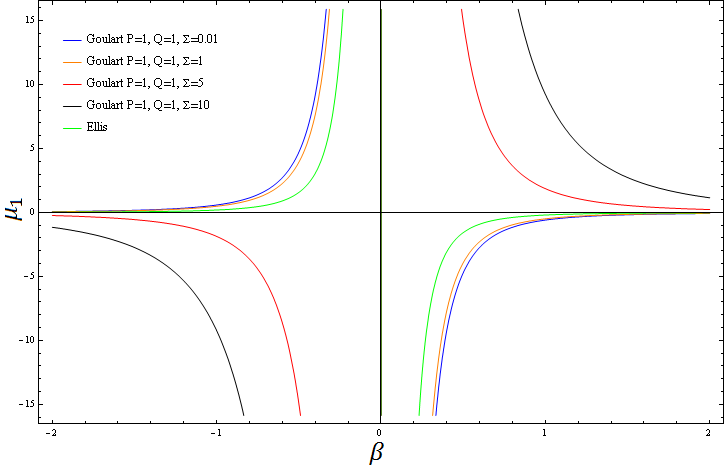}}
\caption{The first-order correction $\protect\mu _{1}$ for the source
angular position $\protect\beta \in $ $[-2;2]$.}
\end{figure}

\begin{figure}[h]
\centerline { \includegraphics 
[scale=0.6]{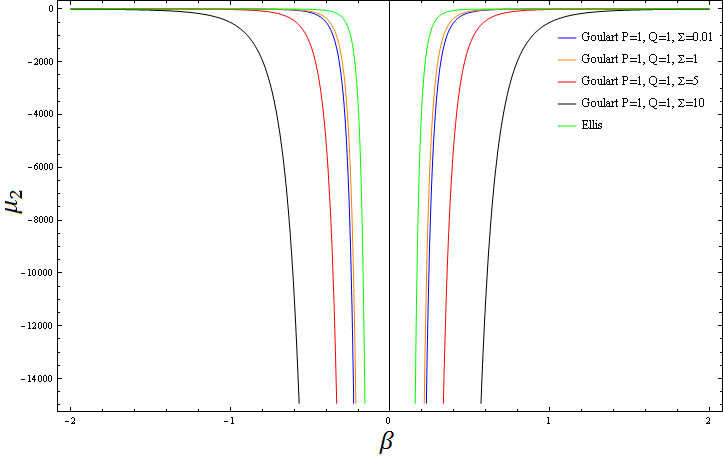}}
\caption{The second-order correction $\protect\mu _{2}$ for the source
angular position $\protect\beta \in $ $[-2;2]$.}
\end{figure}

\begin{figure}[h]
\centerline { \includegraphics 
[scale=0.6]{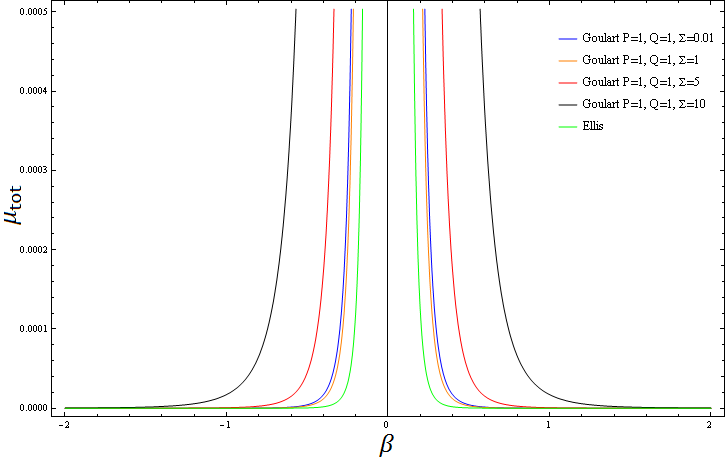}}
\caption{The total magnification $\protect\mu _{\text{tot}}$ for the source
angular position $\protect\beta \in $ $[-2;2]$.}
\end{figure}

\begin{figure}[h]
\centerline { \includegraphics 
[scale=0.6]{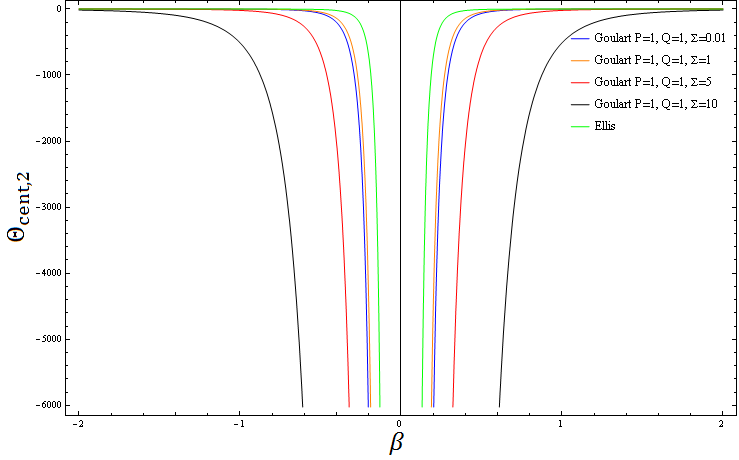}}
\caption{The second-order correction $\Theta _{\text{cent},2}$ for the
source angular position $\protect\beta \in $ $[-2;2]$.}
\end{figure}

\textbf{Acknowledment}

The authors thank an anonymous referee for his/her useful suggestions that
helped improve the paper. The reported study was funded by RFBR according to
the research Project No. 18-32-00377.

\textbf{References}

[1] A. Giahi-Saravani and B. M. Sch\"{a}fer, Mon. Not. R. Astron. Soc. 
\textbf{437}, 2 (2014).

[2] L. Chetouani and G. Clement, Gen. Rel. Grav. \textbf{16,} 111 (1984).

[3] Y. Toki, T. Kitamura, H. Asada and F. Abe, Astrophys. J. \textbf{740},
121 (2011).

[4] V. Perlick, Phys. Rev. D \textbf{69}, 064017 (2004).

[5] K. K. Nandi, Y.Z. Zhang and A. V. Zakharov, Phys. Rev. D \textbf{74},
024020 (2006).

[6] A. Bhattacharya and A.A. Potapov, Mod. Phys. Lett. A \textbf{25}, 2399
(2010).

[7] K. K. Nandi, R. N. Izmailov, A. A. Yanbekov and A. A. Shayakhmetov,
Phys. Rev. D \textbf{95}, 104011 (2017).

[8] A. Tamang, .A A. Potapov, R. Lukmanova, R. Izmailov and K. K. Nandi,
Class.Quant.Grav. \textbf{32}, 235028 (2015).

[9] A. Bhadra, Phys. Rev. D \textbf{67}, 103009 (2003).

[10] T. Harko and F. S. N. Lobo, Phys. Rev. D \textbf{92}, 043011 (2015).

[11] A. Bhadra, K. Sarkar and K. K. Nandi, Phys. Rev. D \textbf{75, }123004
(2007).

[12] C. Cattani, M. Scalia, E. Laserra, I. Bochicchio and K.K. Nandi, Phys.
Rev. D \textbf{87}, 047503 (2013).

[13] C. Armend\'{a}riz-P\'{\i}con, Phys. Rev. D \textbf{65}, 104010 (2002).

[14] K. K. Nandi, A. A. Potapov, R. N. Izmailov, A. Tamang and J. C. Evans,
Phys. Rev. D\textbf{\ 93,} 104044 (2016).

[15] I. D. Novikov and A. A. Shatskiy, J. Exp. Theor. Phys. \textbf{114},
801 (2012).

[16] K. Jusufi, A. \"{O}vgun and A. Banerjee,\ Phys. Rev. D \textbf{96,}
084036 (2017).

[17] M. Visser, Lorentzian Wormholes-From Einstein To Hawking (AIP, New
York, 1995).

[18] K. Jusufi, F. Rahaman and A. Banerjee, Ann. Phys. (Amsterdam)\textbf{389, }219
(2018).

[19] A. Ishihara, Y. Suzuki, T. Ono, T. Kitamura and H. Asada, Phys. Rev. D 
\textbf{94, }084015 (2016).

[20] H. Arakida, Gen. Relativ. Gravit. \textbf{50,} 48 (2018).

[21] N. Tsukamoto, T. Kitamura, K. Nakajima and H. Asada, Phys. Rev. D 
\textbf{90}, 064043 (2014).

[22] K. Jusufi, N. Sarkar, F. Rahaman, A. Banerjee and S. Hansraj,
Eur.Phys. J. C \textbf{78}, 349 (2108).

[23] H. G. Ellis, J. Math. Phys.\textbf{14}, 104 (1973); \textbf{15}, 520
(E) (1974).

[24] K. A. Bronnikov, Acta Phys. Pol. B \textbf{4}, 251 (1973).

[25] F. Abe, Astrophys. J. \textbf{725}, 787 (2010).

[26] R. Lukmanova, A. Kulbakova, R. Izmailov and A. A. Potapov, Int. J.
Theor. Phys. \textbf{55}, 4723 (2016).

[27] P. Goulart, arXiv:1611.03093.

[28] C. R. Keeton and A.O. Petters, Phys. Rev. D\textbf{\ 72,} 104006 (2005).

[29] C. R. Keeton and A. O. Petters, Phys.Rev. D\textbf{73, }044024 (2006).

[30] C. R. Keeton and A. O. Petters, Phys.Rev. D\textbf{73, }104032 (2006).

[31] P. Goulart, Class. Quantum Grav. \textbf{35}, 025012 (2018).

[32] V. Bozza, Phys. Rev. D \textbf{66}, 103001 (2002).

[33] S.V. Iyer and A.O. Petters, Gen. Rel. Grav. \textbf{39}, 1563 (2007).

[34] K.S. Virbhadra, D. Narasimha and S.M. Chitre, Astron. Astrophys. \textbf{337}, 1--8 (1998).

[35] K.S. Virbhadra and, G.F.R. Ellis, Phys. Rev. D \textbf{62}, 084003 (2000).

\bigskip

\end{document}